\documentclass[runningheads]{llncs}
\usepackage{graphicx}
\usepackage{xcolor}
\usepackage{subfigure}
\usepackage{placeins}

\usepackage{bbding}
\usepackage[misc]{ifsym}

\begin{document}
\title{Detecting when pre-trained nnU-Net models fail silently for Covid-19 lung lesion segmentation\thanks{Supported by the Bundesministerium für Gesundheit (BMG) with grant [ZMVI1-2520DAT03A]. The final authenticated version of this manuscript will be published in Lecture Notes in Computer Science, Medical Image Computing and Computer Assisted Intervention - MICCAI 2021 (doi will follow).}}
\titlerunning{Detecting when pre-trained nnU-Net models fail silently for Covid-19}
%
\author{Camila Gonzalez\inst{1}\orcidID{0000-0002-4510-7309}\textsuperscript{(\Letter)}
\and Karol Gotkowski\inst{1}
\and Andreas Bucher\inst{2}
\and Ricarda Fischbach\inst{2}
\and Isabel Kaltenborn\inst{2}
\and Anirban Mukhopadhyay\inst{1}}


%
\authorrunning{C. Gonzalez et al.}
%
\institute{Darmstadt University of Technology, Karolinenpl. 5, 64289 Darmstadt, Germany  \and
University Hospital Frankfurt, Theodor-Stern-Kai 7, 60590 Frankfurt am Main, Germany}


%
\maketitle              
\begin{abstract}

Automatic segmentation of lung lesions in computer tomography has the potential to ease the burden of clinicians during the Covid-19 pandemic. Yet predictive deep learning models are not trusted in the clinical routine due to \emph{failing silently} in out-of-distribution (OOD) data. We propose a lightweight OOD detection method that exploits the Mahalanobis distance in the feature space. The proposed approach can be seamlessly integrated into state-of-the-art segmentation pipelines without requiring changes in model architecture or training procedure, and can therefore be used to assess the suitability of pre-trained models to new data. We validate our method with a patch-based nnU-Net architecture trained with a multi-institutional dataset and find that it effectively detects samples that the model segments incorrectly.

\end{abstract}

\section{Introduction}

Automatic lung lesion segmentation in the clinical routine would significantly lessen the burden of radiologists, standardise quantification and staging of Covid-19 as well as open the way for a more effective utilisation of hospital resources. With this hope, several initiatives have gathered Computed Axial Tomography (CAT) scans and ground-truth annotations from expert thorax radiologists and released them to the public \cite{challenge_data,ma_jun_2020_3757476,morozov2020mosmeddata}. Experts have identified ground glass opacities (GGOs) and consolidations as characteristic of a pulmonary infection onset by the SARS-CoV-2 virus \cite{parekh2020review}. Deep learning models have shown good performance in segmenting these lesions. Particularly the fully-automatic \textit{nnU-Net} framework \cite{isensee2021nnu} secured top spots (9 out of 10, including the first) in the leaderboard for the \textit{Covid-19 Lung CT Lesion Segmentation Challenge} \cite{challenge_results}.

Such frameworks would ideally be utilised in the clinical practice. However, deep learning models are known to fail for data that considerably diverges from the training distribution. CAT scans are particularly prone to this \textit{domain shift} problem \cite{glocker2019machine}. The data showcased in the challenge is multi-centre and diverse in terms of patient group and acquisition protocol. A model trained with it would be presumed to produce good predictions for a wide spectrum of institutions. Yet when we evaluate a nnU-Net model on three other datasets, we notice a considerable drop in segmentation quality (see Fig.~\ref{fig:mahalanobis_merge} (a)). Lung lesions do not manifest in large connected components (see Fig~\ref{fig:qual}), so it is not trivial for a novice radiologist to identify an incorrect segmentation.

Clinicians can still leverage models trained with large amounts of heterogeneous data, but only alongside a process that identifies when the model is unsuitable for a new data sample. Widely-used segmentation frameworks \emph{are not designed with OOD detection in mind}, and so a method is needed that reliably identifies OOD samples post-training while requiring minimal intervention.

Several strategies have shown good OOD detection performance in classification models. Hendrycks and Gimpel \cite{hendrycks2016baseline} propose using the maximum softmax output as an OOD detection baseline. Guo et al. \cite{guo2017calibration} find that replacing the regular softmax function with a \textit{temperature-scaled} variant produces truer estimates. This can be complemented by adding perturbations to the network inputs \cite{liang2018enhancing}. Other methods \cite{hendrycks2019using,lee2018training} instead look at the KL divergence of softmaxed outputs from the uniform distribution. Some approaches use OOD data during training to explicitly train an outlier detector \cite{bevandic2019simultaneous,hendrycks2018deep,lee2018training}. Bayesian-inspired techniques can also be used for outlier detection. Commonly-used are Monte Carlo Dropout \cite{gal2016dropout} and Deep Ensembles \cite{lakshminarayanan2017simple}. These have shown promising results in the field of medical image segmentation \cite{jungo2020analyzing,jungo2019assessing,mehrtash2020confidence}. Approaches that modify the architecture or training procedure have shown better performance in some cases, but their applicability to widely-used segmentation frameworks is limited \cite{blundell2015weight,kohl2018probabilistic,NEURIPS2020_95f8d990}.

We propose a method for OOD detection that is lightweight and seamlessly integrates into complex segmentation frameworks. Inspired by the work of Lee et al. \cite{lee2018simple}, our approach estimates a multivariate Gaussian distribution from in-distribution (ID) training samples and utilises the Mahalanobis distance as a measure of uncertainty during inference. We compute the distance in a low-dimensional feature space, and down-sample it further to ensure a computationally inexpensive calculation. We validate our method on a patch-based 3D nnU-Net trained with multi-centre data from the \textit{Covid-19 Lung CT Lesion Segmentation Challenge}. Our evaluation shows that the proposed method can effectively identify OOD samples for which the model produces faulty segmentations, and provides good model calibration estimates. Our contributions are:

\begin{itemize}
    \item The introduction of a lightweight, flexible method for OOD detection that can be integrated into any segmentation framework.
    \item An extension of the nnU-Net framework to provide clinically-relevant uncertainty estimates.
\end{itemize}

\section{Materials and Methods}

We start by summarising the particularities of the nnU-Net framework in Sec.~\ref{sec:nnunet}. In Sec.~\ref{sec:est}, we outline our proposed method for OOD detection, which follows a \emph{three-step process}: (1) estimation of a Gaussian distribution from training features (2) extraction of uncertainty masks for test images and (3) calculation of subject-level uncertainty scores.

\subsection{Patch-based nnU-Net} \label{sec:nnunet}

The nnU-Net framework is a standardised baseline for medical image segmentation \cite{isensee2021nnu}. Without deviating from traditional U-Net architectures \cite{ronneberger2015u}, it has won several grand challenges by automatically customising the architecture and training configuration to the data at hand \cite{challenge_results}. The framework also performs pre- and post-processing steps, such as adapting voxel spacing and contrast normalisation, during both training and inference. In this work we utilise the patch-based full-resolution variant, which is recommended for most applications \cite{isensee2021nnu}, but our method can be integrated into any other architecture. For the patch-based architecture, training images are first divided into overlapping patches with a sliding window approach, resulting in $N$ patches $\left\{ x_i\right\}^{N}_{i=1}$. Predictions for each patch are multiplied by a filtering operation that weights centre-voxels more heavily, and then aggregated into an output mask with the dimensions of the original image.

\begin{figure}
\centering
\includegraphics[width=\textwidth]{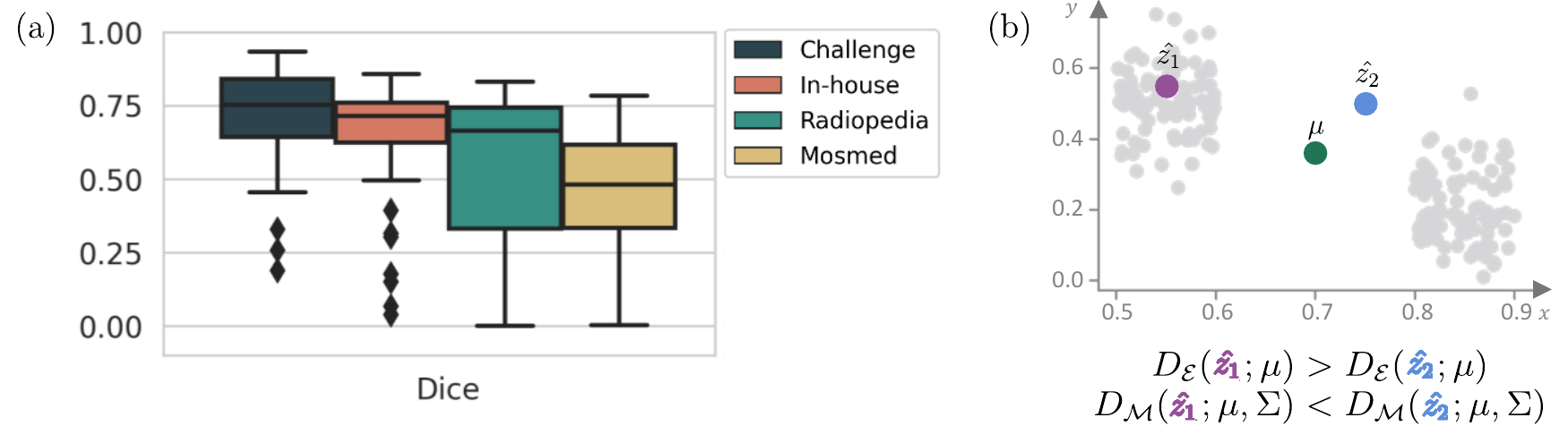}
\caption{(a) Dice coefficient of a model trained with \textit{Challenge} data, evaluated with ID (\textit{Challenge}) test data as well as on three other datasets. (b) The euclidean distance $D_\mathcal{E}$ does not recognize that $\hat{z_1}$ (purple marker) is closer than $\hat{z_2}$ (blue marker) to the distribution of training samples (gray markers), with mean $\mu$ (green marker) and covariance $\Sigma$. This difference intensifies in high-dimensional spaces, where it is common for regions close to the mean to be underrepresented.} \label{fig:mahalanobis_merge}
\end{figure}

\subsection{Estimation of a subject-level uncertainty score} \label{sec:est}

We are interested in capturing \textit{epistemic uncertainty}, which arises from a lack of knowledge about the data-generating process. Quantifying it for image \textit{regions} instead of region boundaries is challenging, particularly for OOD data \cite{kendall2017uncertainties}. One computationally inexpensive way to assess epistemic uncertainty is to calculate the distance between training and testing activations in a low-dimensional feature space. As a model is unlikely to produce reasonable outputs for features far from any seen during training, this is a reliable signal for bad model performance \cite{lee2018simple}. Model activations have covariance and the activations of typical input images do not necessarily resemble the mean \cite{wei2015understanding}, so the euclidean distance is not appropriate to identify unusual activation patterns; a problem that exacerbates in high-dimensional spaces. The Mahalanobis distance $D_{\mathcal{M}}$ rescales samples into a space without covariance, supplying a more effective way to identify typical patterns in deep model features. Fig.~\ref{fig:mahalanobis_merge} (b) illustrates a situation where the euclidean distance assumes that $\hat{z_2}$ is closer to the training distribution than $\hat{z_1}$, when $\hat{z_2}$ is highly unusual and $\hat{z_1}$ is a probable sample.

In the following we describe the steps we perform to extract a subject-level uncertainty value. Note that only one forward pass is necessary for each image, keeping the computational overhead to a minimum.

 \subsubsection{Estimation of the training distribution:} We start by estimating a multivariate Gaussian $\mathcal{N}(\mu, \Sigma)$ over model features. For all training inputs $\left\{ x_i\right\}^{N}_{i=1}$, features $\mathcal{F}(x_i)=z_i$ are extracted from the encoder $\mathcal{F}$ of the pre-trained model. For modern segmentation networks, the dimensionality of the extracted features $z_i$ is too large to calculate the covariance $\Sigma$ in an acceptable time frame. We thus project the latent space into a lower subspace by average pooling. Finally, we flatten this subspace and estimate the empirical mean $\mu$ and covariance $\Sigma$.

\begin{equation} \label{eq:cov}
    \mu = \frac{1}{N} \sum_{i=1}^N\hat{z_i}, \quad  \Sigma = \frac{1}{N} \sum_{i=1}^N (\hat{z_i} - \mu)(\hat{z_i} - \mu)^T
\end{equation}

\subsubsection{Extraction of uncertainty masks:} 
\begin{figure}
\centering
\includegraphics[width=0.9\textwidth]{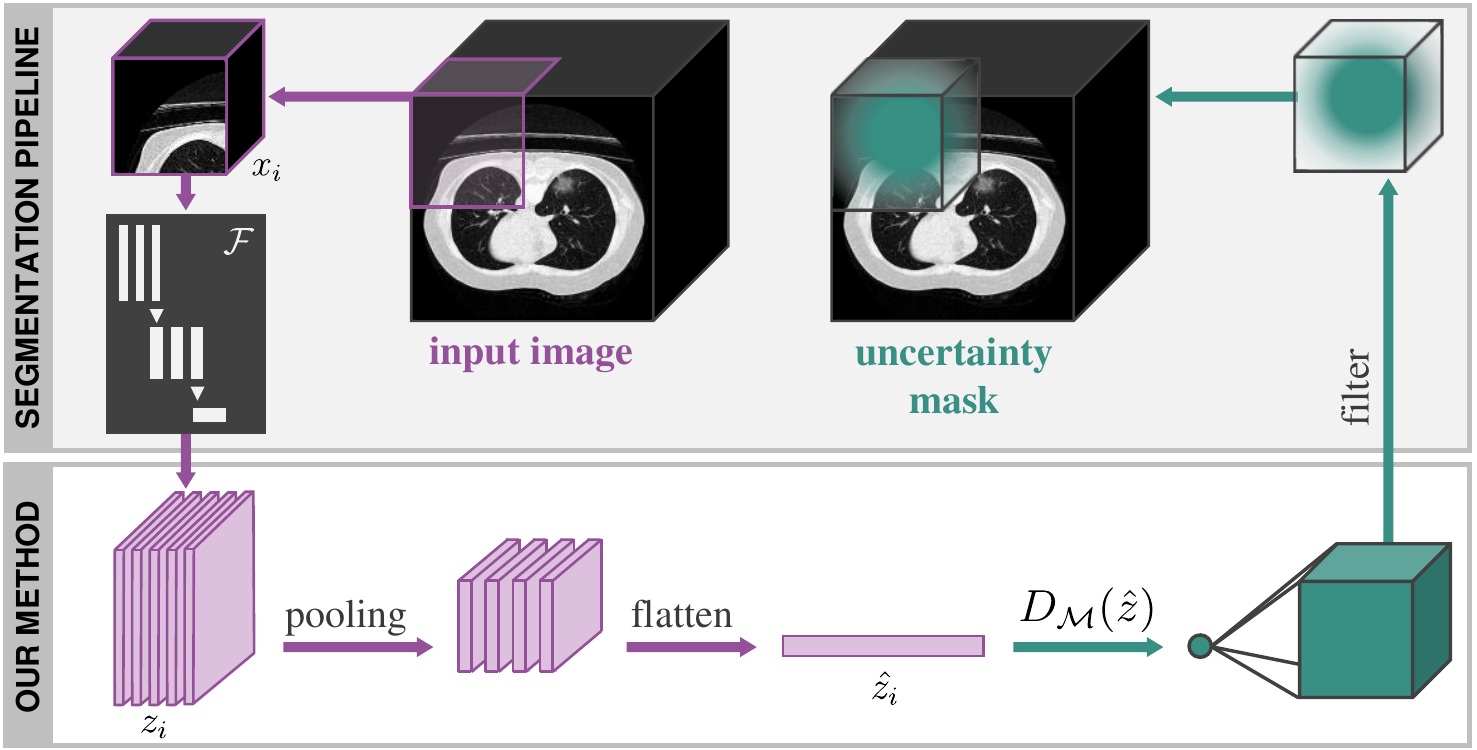}
\caption{Extracting an uncertainty mask based on the Mahalanobis distance $D_\mathcal{M}$ for an image during inference, in combination with a patch-based nnU-Net architecture.
} \label{fig:methodology}
\end{figure}

During inference, we estimate an uncertainty mask for a subject following the process outlined in Fig.~\ref{fig:methodology}. For each patch $x_i$, features are extracted and projected into $\hat{z_i}$. Next, the Mahalanobis distance (Eq.~\ref{eq:mahalanobis}) to the Gaussian distribution estimated in the previous step is calculated.

\begin{equation} \label{eq:mahalanobis}
    D_\mathcal{M}(\hat{z_i};\mu, \Sigma) = (\hat{z_i} - \mu)^T \Sigma^{-1}(\hat{z_i} - \mu)
\end{equation}

Each distance is a point estimate for the corresponding model input. These are aggregated in a similar fashion to how network outputs are combined to form a prediction mask. Following the example of the patch-based nnU-Net, a zero-filled tensor is initialised with the dimensionality of the original image. After assessing the distance for a patch, the value is replicated to the specified patch size and a filtering operation is applied to weight centre voxels more heavily. Finally, patch-level uncertainties are aggregated to an image-level mask.

\subsubsection{Subject-level uncertainty:} The process described above produces an uncertainty mask with the dimensionality of the CAT scan. In order to effectively identify highly uncertain samples, we aggregate these into a subject-level uncertainty $\mathcal{U}$ by averaging over all voxels. We then normalise uncertainties between the minimum and doubled maximum values represented in an ID validation set -- which we assume to be available during training -- to ensure $\mathcal{U}\in \left[ 0, 1 \right]$.

\section{Experimental Setup}

We work with a total of four datasets for segmentation of Covid-19-related findings. The \textit{Challenge} dataset \cite{challenge_data} contains chest CAT scans for patients with a confirmed SARS-CoV-2 infection from an array of institutions. The data is heterogeneous in terms of age, gender and disease severity. We use the 199 cases made available under the \textit{Covid Segmentation Grand Challenge}, which we randomly divide into 160 cases to train the model, 4 validation and 35 test cases. 

We evaluate our method with two publicly available datasets and an in-house one. The public datasets encompass cases for patients with and without confirmed infections. \textit{Mosmed} \cite{morozov2020mosmeddata} contains fifty cases and the \textit{Radiopedia} dataset \cite{ma_jun_2020_3757476}, a further twenty. Finally, we utilise an in-house dataset consisting of fifty patients who were tested positive for SARS-CoV-2 with an RT PCR test. All fifty scans were reviewed for diagnostic image quality. The annotations for the in-house data were performed slice-by-slice by two independent readers trained in the delineation of GGOs and pulmonary consolidations. Central vascular structures and central bronchial structures were excluded from all segmentations. All delineations were reviewed by an expert radiologist reader. For the public datasets, the segmentation process is outlined in the corresponding publications.

With the \textit{Challenge} data, we train a patch-based nnU-Net \cite{isensee2021nnu} on a \textit{Tesla T4} GPU. Our configuration has a patch size of $\left [ 28, 256, 256 \right ]$, and adjacent patches overlap by half that size. To reduce the dimensionality of the feature space, we apply average pooling with a kernel size of $\left ( 2,2,2 \right )$ and stride $\left ( 2,2,2 \right )$ until the dimensionality falls below $1e4$ elements. With the \textit{Scikit Learn} library (version 0.24) \cite{scikit_learn}, calculating $\Sigma$ requires 85 seconds for $1e5$ samples. Our code is available under github.com/MECLabTUDA/Lifelong-nnUNet (branch \textit{ood\_detection}).

We compare our approach to state-of-the-art techniques to assess uncertainty information by performing inference on a trained model. \emph{Max. Softmax} consists of taking the maximum softmax output \cite{hendrycks2016baseline}. \emph{Temp. Scaling} performs temperature scaling on the outputs before applying the softmax operation \cite{guo2017calibration}, for which we test three different temperatures $T=\left \{ 10, 100, 1000 \right \}$. \emph{KL from Uniform} computes the KL divergence from an uniform distribution \cite{hendrycks2019using}. Note that all three methods output a \textit{confidence} score (higher is more certain), which we invert to obtain an \textit{uncertainty} estimate (lower is more certain). Finally, \emph{MC Dropout} consists of doing several forward passes whilst activating the Dropout layers that would usually be dormant during inference. We perform $10$ forward passes and report the standard deviation between outputs as an uncertainty score. For all methods, we calculate a subject-level metric by averaging uncertainty masks, and normalise the uncertainty range between the minimum and doubled maximum uncertainty represented in ID validation data.

\section{Results}

We start this section by analysing the performance of the proposed method in detecting samples that vary significantly from the training distribution. We then examine how well the model estimates segmentation performance. Lastly, we qualitatively evaluate our method for ID and OOD examples.

\subsubsection{OOD detection:} We first assess how effective our method is at identifying samples that are not ID (\textit{Challenge} data). Due to the heterogeneity of the \textit{Challenge} dataset, in practice data from an array of institutions would be considered ID. However, for our evaluation datasets there is a drop in performance which should manifest in higher uncertainty estimates. As is common practice in OOD detection \cite{liang2018enhancing}, we find the uncertainty boundary that achieves a $95\%$ true positive rate (TPR) on the ID validation set, where a \textit{true positive} is a sample correctly identified as ID. We report for the ID test data and all OOD data the false positive rate (FPR) and \textit{Detection Error} $\ =0.5\ (1-TPR)+0.5\ FPR$ at $95\%$ TPR. Tab.~\ref{tab:ood_detection} summarizes our findings. All methods that utilise the network outputs after one forward pass have a high detection error and FPR, while the MC Dropout approach manages to identify more OOD samples. Our proposed method displays the lowest FPR and detection error.

\begin{table}
\centering
\caption{Detection Error (lower is better) and FPR (lower is better) for the boundary of 95\% TPR, ESCE (lower is better) and (mean±sd) Dice (higher is better) for subjects with an uncertainty below the 95\% TPR boundary. The results are reported for ID test data and all OOD samples.}\label{tab:ood_detection}
\begin{tabular}{p{4cm}|p{2cm}|p{1.5cm}|p{1.5cm}|p{2.2cm}}

\textbf{Method} & \textbf{Det. Error} & \textbf{FPR} & \textbf{ESCE} & \textbf{Dice} \\
\hline
\hline
Max. Softmax \cite{hendrycks2016baseline} & 0.334 & 0.583& 0.319  & 0.582 \textpm 0.223\\
Temp. Scaling $T=10$ \cite{guo2017calibration} & 0.508 & 0.758 & 0.407 & 0.601 \textpm 0.233\\
Temp. Scaling $T=100$ \cite{guo2017calibration} & 0.361 & 0.550 & 0.408 & 0.589 \textpm 0.233\\
Temp. Scaling $T=1000$ \cite{guo2017calibration} & 0.500 & 1.000 & 0.408 & 0.592 \textpm 0.233\\
KL from Uniform \cite{hendrycks2019using} & 0.415 & 0.717& 0.288 & 0.600 \textpm 0.215 \\
MC Dropout \cite{gal2016dropout} & 0.177 & 0.183 & 0.215 & 0.614 \textpm 0.234\\
\textbf{Ours} & \textbf{0.082} & \textbf{0.050} & \textbf{0.125} & \textbf{0.744 \textpm 0.143}\\

\end{tabular}
\end{table}

\textbf{Segmentation performance:} While the detection of OOD samples is a first step in assessing the suitability of a model, an ideal uncertainty metric would inversely correlate with model performance, informing the user of the likely quality of a prediction without requiring manual annotations. For this we calculate the \textit{Expected Segmentation Calibration Error} (ESCE). Inspired by Guo et al. \cite{guo2017calibration}, we divide the $N$ test scans into $M=10$ interval bins $B_m$ according to their normalised uncertainty. Over all bins, the absolute difference is added between average Dice $(Dice(B_m))$ and inverse average uncertainty $(1 - \mathcal{U}(B_m))$ for samples in the bin, weighted by the number of samples. 

\begin{equation} \label{eq:ece}
    ESCE=\sum ^M_{m=1}\frac{\left | B_m \right |}{N}\left | Dice(B_m)-(1-\mathcal{U}(B_m)) \right |
\end{equation}

The results are reported in Tab.~\ref{tab:ood_detection} (forth column). Our proposed approach shows the lowest $ESCE$ at 0.125. The average Dice of admitted samples (fifth column) lies at 0.744, which is consistent with the ID expected performance of the model (see Fig.~\ref{fig:mahalanobis_merge} (a)).

\begin{figure}
\centering
\includegraphics[width=0.9\textwidth]{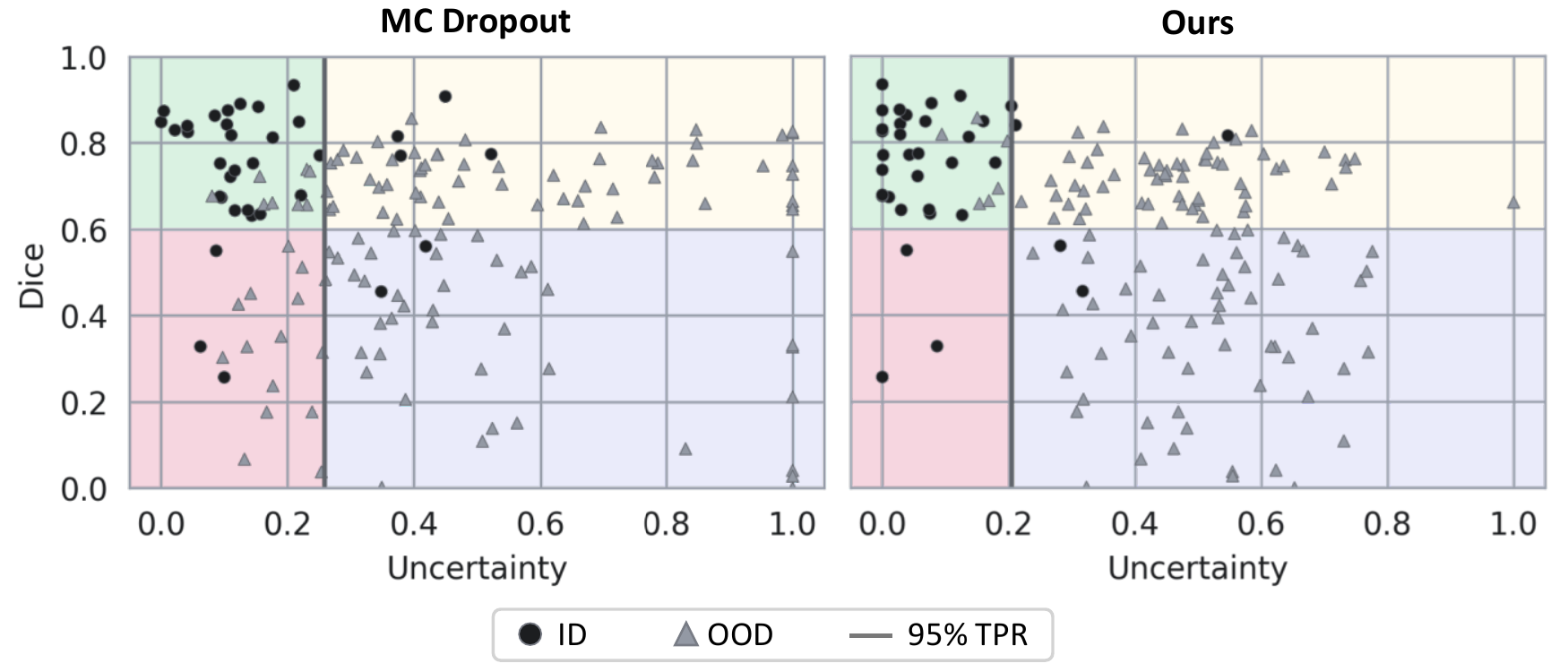}
\caption{Dice coefficient against normalised uncertainty for OOD (gray triangles) and test ID (black circles) samples. The vertical gray line marks the boundary of 95\% TPR for ID validation data. To the right of this line, samples are classified as OOD. The lower left (red) quadrant is clinically most relevant. Unlike MC Dropout, our method does not fail silently by assigning low uncertainties to low-Dice samples.} \label{fig:scatters}
\end{figure}

Fig.~\ref{fig:scatters} depicts the Dice coefficient plotted against the uncertainty for our proposed approach and MC Dropout, which has the second lowest calibration error. Relevant for a safe use of the model in clinical practice is the lower left (red) quadrant, where \emph{silent failures} are located. Whereas MC Dropout fails to identify OOD samples with faulty predicted segmentations, our proposed method assigns these cases high uncertainty estimates. The only OOD samples that fall below the 95\% TPR uncertainty boundary for our method have Dice scores over 0.6 (upper left quadrant with green background). However, our method shows room for improvement in the upper right (yellow) quadrant. Here, OOD samples for which the model produces good predictions are estimated to have a high uncertainty. An ideal calibration would place all samples in the upper left (green) and lower right (blue) quadrants.

\begin{figure}
\centering
\includegraphics[width=\textwidth]{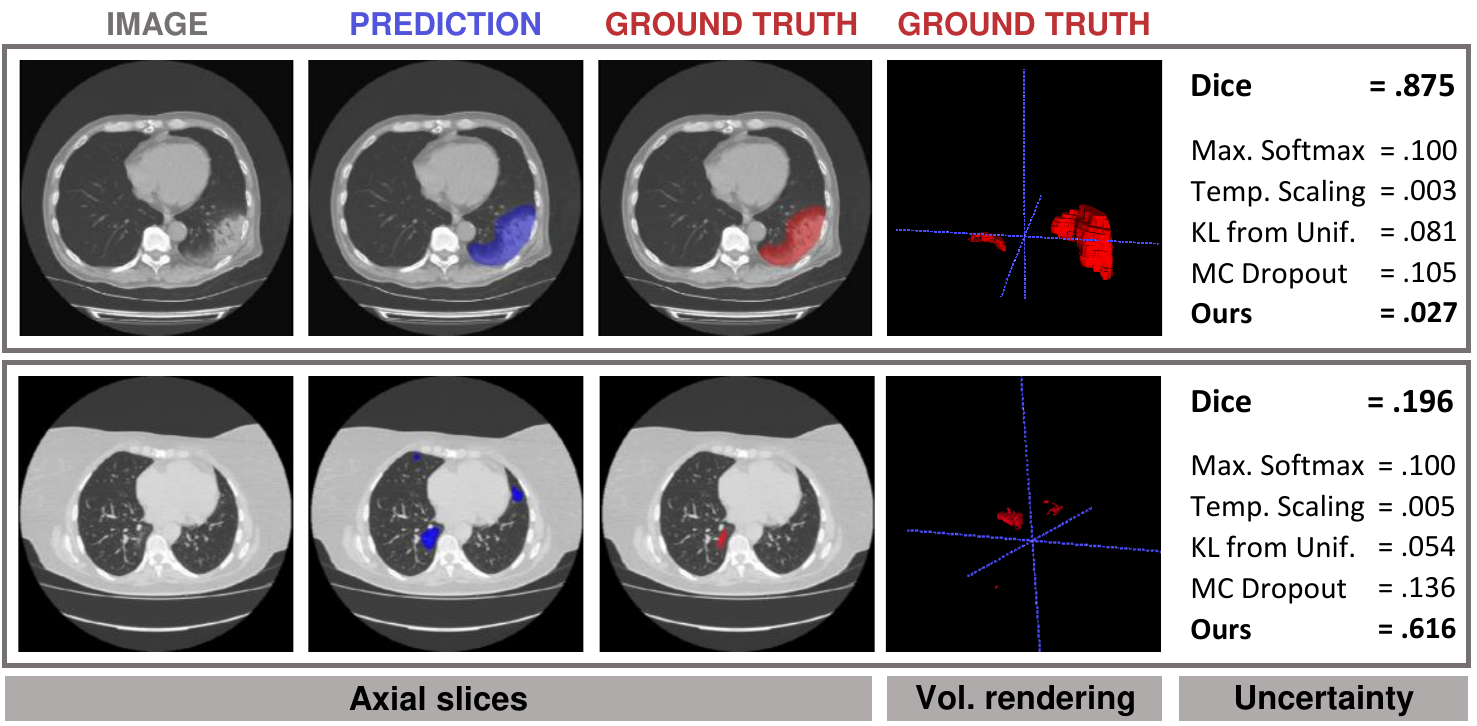}
\caption{Upper row: a good prediction. Lower row: a prediction for an OOD sample where two lesions are erroneously segmented in the superior lung lobes. Despite the considerable differences to the ground truth, these errors are not directly noticeable for the inexpert observer, as GGOs can manifest in superior lobes \cite{parekh2020review}.} \label{fig:qual}
\end{figure}

\textbf{Qualitative evaluation:} Fig.~\ref{fig:qual} depicts two example images alongside corresponding ground truths and predictions. The top row shows a example from the \textit{Challenge} dataset for which the model produces an adequate segmentation. The bottom contains a scan from the \textit{Mosmed} dataset. The model oversegments the lesion at the middle left lobe and incorrectly marks two additional regions at the left and right superior lobes. Only our proposed method signals a possible error in the lower row with a high uncertainty, while producing a low uncertainty estimate for the upper row.

\FloatBarrier
\section{Conclusion}

Increasingly, institutions are taking part in initiatives to gather large amounts of annotated, heterogeneous data and release it to the public. This could potentially alleviate the work burden of medical practitioners by allowing the training of robust segmentation models. Open-source end-to-end frameworks contribute to this process. But regardless of the variety of the training data, it is necessary to assess whether a model is well-suited to new samples. This is particularly true when it is not trivial to identify a faulty output, such as for the segmentation of SARS-CoV-2 lung lesions. There is currently a disconnect between methods for OOD detection, which often require special training or architectural considerations, and widely-used segmentation frameworks. We find that calculating the Mahalanobis distance to features in a low-dimensional subspace is a lightweight and flexible way to signal when a model prediction should not be trusted. Future work should explore how to better identify high-quality predictions. For now, our work increases clinicians' trust while translating trained neural networks from challenge participation to real clinics.

%
%
%
\newpage
\bibliographystyle{splncs04}
\bibliography{paper1378}

\end{document}